\documentclass[trackchanges]{aastex701}

\usepackage[utf8]{inputenc}
\usepackage{amsmath}
\usepackage{graphicx}

\usepackage{xcolor}
\usepackage{natbib}
\hypersetup{ colorlinks,linkcolor={blue!80!black},citecolor={blue!50!black}, urlcolor={blue!80!black}
}
\usepackage[format=plain, labelfont={bf}, ]{caption}

\usepackage [english]{babel}
\usepackage [autostyle, english = american]{csquotes}
\usepackage{hyperref}
\MakeOuterQuote{"}

\begin{document}

\title{Emulation of the Halo Mass Function with Neural Networks}

\author[orcid=0000-0001-6778-6659,gname=Olive, sname='Ross']{Olive Ross}
\affiliation{Cornell University}
\email[show]{ogr8@cornell.edu}  

\author[orcid=0000-0001-5846-0411,gname=Nicholas, sname='Battaglia']{Nicholas Battaglia} 
\affiliation{Cornell University}
\email{nbatta@astro.cornell.edu}

\begin{abstract}

The abundance of galaxy clusters as a function of mass --- the halo mass function (HMF) --- depends strongly on cosmological parameters. Large-scale simulations are computationally expensive, motivating emulators that interpolate simulation outputs. We present a neural network emulator that maps cosmological and astrophysical parameters, halo mass, and redshift directly to a cumulative count of halos above the input mass, avoiding the information loss inherent to binning and to the functional or PCA-based data reduction used in prior emulators. Each emulator prediction is accompanied by a self-estimated error, expressed in units of Poisson shot noise, that the network learns to predict alongside the halo count itself — giving every output a built-in confidence metric. We validate the emulator with leave-one-out tests on three simulation suites --- the N-body MassiveNus suite and the hydrodynamical CAMELS IllustrisTNG and SIMBA suites --- spanning a range of box sizes, redshifts, sub-grid physics, and softwares. Of all the held-out predictions ($1.2\times10^6$ in total), 88\% have model bias at or below shot noise and 99.9\% are within $3\times$ shot noise. The only significant high-error tail is in MassiveNus, where the predicted model bias flags 76\% of predictions exceeding $3\times$ shot noise; the remaining number of high error predictions is comparable to the number expected from shot noise alone. This emulator is publicly available and can be trained on other simulation suites making it a flexible HMF tool that can be used for cosmological and astrophysical inference.

\end{abstract}

\section{Introduction}
\label{sec:intro}

The evolution of cosmic structure is sensitively dependent on a wide array of cosmological and astrophysical processes. From primordial density fluctuations to feedback from active galactic nuclei, the late-time structure we observe today is encoded with a wealth of information. By quantifying the effects of these processes on structure growth, we can further our understanding of the origins and evolution of our universe.

Galaxy clusters---the most massive gravitationally collapsed structures in the universe---trace the filaments of large-scale structure, giving insight into the statistics of structure formation (\citet{haiman_constraints_2001, hu_sample_2003, mantz_weighing_2015}; see \citet{allen_cosmological_2011} for a review). The distribution of cluster masses is sensitive to cosmological and astrophysical parameters, so large-scale cosmological simulations are run to find the distribution of halos as a function of mass under different conditions. These simulation results are commonly condensed into analytic fitting functions that describe how the HMF depends on redshift, halo definition, and cosmology; the ansatz of \citet{tinker_toward_2008}, calibrated against a large suite of N-body simulations, remains one of the most widely used and is adopted for comparison in this work (Section ~\ref{sec: Results}).

Numerous surveys have constructed large cluster catalogs through observations of their optical light (e.g., SDSS \citet{wen_catalog_2012}, DES \citet{rykoff_redmapper_2016}), X-ray emissions (CHANDRA \citet{vikhlinin_chandra_2009}, eROSITA \citet{bulbul_srgerosita_2024}, ROSAT \citet{bohringer_extended_2013}), and the Sunyaev-Zel'dovich effect (ACT \citet{hasselfield_atacama_2013}, SPT \citet{bocquet_cluster_2019}, Planck \citet{hilton_atacama_2021}). Next-generation surveys now underway --- including Rubin/LSST \citep{ivezic_lsst_2019}, Euclid \citep{laureijs_euclid_2011}, and the Simons Observatory \citep{ade_simons_2019} --- along with future facilities such as CCAT \citep{ccat-prime_collaboration_ccat-prime_2023}, the Nancy Grace Roman Space Telescope \citep{spergel_wide-field_2015}, and NewAthena \citep{valotti_cosmological_2018}, will increase cluster samples by orders of magnitude, placing ever-tighter demands on the accuracy of theoretical HMF predictions.

The number density of clusters as a function of mass is referred to as the halo mass function (HMF), where a halo refers to an overdensity of mass compared to the cosmological background. Halo mass, however, is not directly observable: cluster masses are instead inferred from survey observables such as optical richness, X-ray luminosity, or SZ signal via a mass--observable relation $P(\Theta|M)$, which gives the probability that a halo of mass $M$ produces observable $\Theta$ (e.g.,\citealt{johnston_cross-correlation_2007, maughan_lx-yx_2007, sheldon_cross-correlation_2009}). This relation allows us to forward model a parameter-dependent distribution of observables $n(\Theta|\Psi)$ as follows :

\begin{equation}
    \label{eq:obsdist}
    n(\Theta|\Psi) = \int P(\Theta|M)\dfrac{dn(M|\Psi)}{d\ln(M)}d\ln(M),
\end{equation}
where $\Psi$ is a set of cosmological and/or astrophysical parameters, $M$ is mass, $\Theta$ represents an observable, $P(\Theta|M)$ is the mass-observable relation, $n(M)$ is the number density of halos with masses larger than $M$, and $\dfrac{dn(M)}{d\ln(M)}$ is the HMF \citep{allen_cosmological_2011}. By comparing forward-modeled observable distributions to observations, constraints can be placed on cosmological and astrophysical parameters (e.g., \citet{mantz_weighing_2015, ghirardini_srgerosita_2024, bocquet_halo_2016, bocquet2024sptclustersdeshst}). The precision of these constraints is ultimately limited by the accuracy of the HMF, motivating the search for accurate and generalizable mappings from cosmological and astrophysical parameters to halo mass distributions.

Early models of the HMF advocated for a universal form --- meaning the shape of the distribution depends solely on the matter density fluctuations in the early universe when expressed in appropriate variables. \citet{press_formation_1974} developed the foundational analytical framework for this prescription, which was subsequently refined through adjustments to the multiplicity function, mass barrier, and mass definition as larger simulations became feasible (e.g., \citet{bond_excursion_1991, sheth_ellipsoidal_2001, jenkins_mass_2001, warren_precision_2006}). However, further work demonstrated that the HMF is inherently non-universal, with dependence on redshift \citep{reed_evolution_2003, tinker_toward_2008, bhattacharya_mass_2011}, cosmological parameters such as the dark energy equation of state \citep{haiman_constraints_2001, bhattacharya_mass_2011}, baryonic feedback \citep{bocquet_halo_2016}, and neutrino mass \citep{ryu_constraining_2020}. In particular, \citet{tinker_toward_2008} provided a widely adopted ansatz in which the multiplicity function's parameters are calibrated as explicit functions of the spherical-overdensity threshold $\Delta$ and redshift, capturing much of this non-universality while retaining a simple functional form. Universal fitting functions achieve $\mathcal{O}(5\%)$ accuracy \citep{white_mass_2002}, but this is insufficient for precision cosmology with next-generation surveys. This work explicitly leverages the non-universality of the HMF, fitting a general mapping from cosmological and astrophysical parameters to halo mass distributions rather than prescribing a functional form.

Emulators find patterns between simulation inputs and outputs, enabling predictions for parameter combinations that were never explicitly simulated and allowing fine-grained exploration of parameter space without the cost of additional simulations. The most common approach in cosmology uses Gaussian processes, first proposed for this purpose by \citet{heitmann_cosmic_2006} and \citet{habib_cosmic_2007}. In this approach, the target function (e.g., the HMF) is first fit to a parametric form, whose parameters are then dimensionally reduced to principal component weights via PCA. Gaussian processes are then used to map simulation input parameters to these PCA weights, producing a function that can reproduce simulation results for any set of input parameters.

Variations on this approach have been used to model galaxy clustering \citep{zhai_aemulus_2019}, the galaxy power spectrum \citep{kwan_cosmic_2015}, the concentration-mass relation \citep{kwan_cosmic_2013}, and the matter power spectrum (e.g.,\citet{heitmann_coyote_2009, lawrence_coyote_2010, heitmann_coyote_2013, heitmann_mira-titan_2016, lawrence_mira-titan_2017}). More recently, this method has been applied to the halo mass function by \citet{mcclintock_aemulus_2019}, \citet{nishimichi_dark_2019}, and \citet{bocquet_mira-titan_2020}, who have constructed HMF emulators with percent-level accuracy for masses up to $\sim10^{15}\,M_\odot$ and redshifts up to $z=2$.

These GP-based emulators represent a significant advance, but carry limitations inherited from their design. Fitting the HMF to a parametric function before applying PCA and Gaussian processes constrains predictions to a narrow functional form, and the pipeline must be substantially redesigned for each new simulation suite --- the parametric form, PCA basis, and GP training are all specific to the parameters and outputs of a given set of simulations. As next-generation surveys such as Rubin/LSST and Euclid push mass calibration uncertainties toward the $\sim$1\% level \citep{mcclintock_aemulus_2019}, there is a need for emulation methods that are both more flexible in functional form and general enough to be applied to any simulation campaign without reconfiguration.

To allow for a more flexible and generalizable model, we use neural networks. Neural networks have been gaining in popularity as a method of representing highly complex functions, especially those mapping multiple inputs to multiple outputs. One of the first neural network emulators in cosmology was pkANN \citep{agarwal_pkann_2012, agarwal_pkann_2014}, which modeled the matter power spectrum. More recently this approach has been used to emulate the redshift space power spectrum \citep{kobayashi_accurate_2020}, galaxy power spectrum \citep{donald-mccann_matryoshka_2022}, galaxy correlation function \citep{cuesta-lazaro_galaxy_2022}, and concentration-mass relation \citep{ruan_emulator-based_2023}. \citet{cuesta-lazaro_galaxy_2022} and \citet{ruan_emulator-based_2023} briefly explore applications to the HMF.

In this work, we use mass, redshift, and various cosmological/astrophysical parameters as our inputs, and use a neural network to map these to an expected count of clusters above the input mass. In addition to predicting halo counts, the network simultaneously outputs a self-calibrating uncertainty $\sigma_p$ that estimates the model bias for any input cosmology, providing a built-in reliability metric when emulating outside the training set. This approach avoids data reduction such as parametric prescriptions of the halo mass function, binning, or reduced dimension PCA analysis, and we demonstrate that predictions achieve model bias at or below shot noise across a range of simulation suites, mass scales, and redshifts. In Section~\ref{sec: Simulations and Data} we present the simulations and data used; Section~\ref{sec: The Neural Network} describes the details of the emulator. Results are presented in Section~\ref{sec: Results}; conclusions and future work are discussed in Section~\ref{sec: Discussion and Conclusion}.

\section{Simulations and Data}
\label{sec: Simulations and Data}
\begin{table}

    \begin{tabular}{|c|c|c|c|c|c|c|}
        \hline
             & $\Omega_M$ & $\sigma_8$ & $\mathrm{SN}_1$, $\mathrm{SN}_2$, $\mathrm{AGN}_1$, and $\mathrm{AGN}_2$ & $n_s$ & $\Omega_b$ \\
             \hline
             Fiducial & 0.3 & 0.8 & 1.0 & 0.9624 & 0.049\\
             \hline
             Range  & 0.1 - 0.5 & 0.6 - 1.0 & 0.25 - 4.0 & fixed & fixed \\
             \hline
        \end{tabular}
        \caption{Fiducial values and ranges for CAMELS (IllustrisTNG and SIMBA) simulation parameters.}

        \label{tab:CAMELSfid}
\end{table}
\begin{table}

    \begin{tabular}{|c|c|c|c|c|c|}
    \hline
          & $M_\nu$ & $\Omega_M$ & $A_s$ & $n_s$ & $\Omega_b$\\
         \hline
         Fiducial & 0.0 & 0.3 & $2.1\times10^{-9}$ & 0.97 & 0.046 \\
         \hline
         Range  & 0.0 - 0.6 eV & 0.184 - 0.416 & $1.289\times10^{-9}$ - $2.911\times 10^{-9}$ & fixed & fixed \\
         \hline
        
    \end{tabular}
    \caption{Fiducial values and ranges for MassiveNus simulation parameters.}

    \label{tab:MNusfid}
    
\end{table}


We use three simulation suites with varying cosmological and astrophysical parameters: MassiveNus \citep{liu_massivenus_2018}, CAMELS IllustrisTNG, and CAMELS SIMBA \citep{villaescusa-navarro_camels_2023}.  MassiveNus is a suite of N-body simulations that samples 102 different combinations of primordial power spectrum perturbations  $A_s$, matter densities $\Omega_M$, and neutrino masses $M_\nu$. 
The simulations are evolved forward from $z = 45$ to $z = 0$ and snapshots are taken at 67 redshifts in that range. The CAMELS IllustrisTNG and SIMBA hydrodynamic simulations vary matter density $\Omega_M$, $\sigma_8$, two AGN feedback parameters ($\mathrm{AGN}_1$ and $\mathrm{AGN}_2$), and two supernova feedback parameters ($\mathrm{SN}_1$ and $\mathrm{SN}_2$). $\mathrm{AGN}_1$ is the energy of AGN feedback and $\mathrm{AGN}_2$ is the ejection speed of black hole feedback. $\mathrm{SN}_1$ and $\mathrm{SN}_2$ represent the flux of galactic wind feedback and speed of galactic winds respectively. The specifics of $\mathrm{AGN}_1$ and $\mathrm{SN}_1$ vary between SIMBA and Illustris, see \citet{villaescusa-navarro_camels_2023} for a more detailed description of each parameter. Each CAMELS simulation suite contains 1092 hydrodynamic simulations with 34 snapshots taken between $z = 127$ and $z = 0$. The fiducial values and ranges of parameters for all simulations are shown in Tables ~\ref{tab:CAMELSfid} and ~\ref{tab:MNusfid}.

For each snapshot, halos are identified using the spherical overdensity method \citep{lacey_merger_1993}. In this method, a density coefficient $\Delta$ is chosen; we use $\Delta = 200$ so that our results are minimally sensitive to simulation resolution \citep{tinker_toward_2008}.  The halo's density peak is identified and $R_\Delta$ is increased until the following condition is met:

\begin{equation}
    \label{eq:overdensity}
    \rho_\Delta = \dfrac{M_\Delta}{4/3\pi R_\Delta^3} =\rho_c \Delta, 
\end{equation}
where $\rho_c$ is defined as the critical density at a given redshift and cosmology. We then create a catalog of objects with masses between $10^{13}$ and $10^{15}\,M_\odot$ for the MassiveNus simulations and between $10^{12}$ and $10^{14}\,M_\odot$ for the CAMELS simulations.

\section{The Emulator}
\label{sec: The Neural Network}

\subsection{Cumulative Distributions}
Our goal is to constrain a continuous halo mass function from discrete data (i.e. halo masses). This is challenging because the probability of finding a halo with exactly mass M in a discrete simulation vanishes. The standard solution to this is binning which produces the probability of finding a halo within a range of masses. However, binning erases information within the bin that could be vital for precision models. For this reason, we introduce a cumulative distribution (CDF) of masses as in \citet{ruan_emulator-based_2023} and \citet{Chen_2025}. We create a function that returns the number of simulated halos with mass greater than the input value $M$ 

\begin{equation}
   N( > M; z, C) = \int_M^\infty \mathrm{d}m \dfrac{\mathrm{d}n(m;z, C)}{\mathrm{d}m} \mathrm{d}V.
\end{equation}
Where $z$ is redshift, $C$ is some set of cosmological and/or astrophysical parameters, and $V$ is simulation box size. Note that this is the complement of a CDF in the usual statistical sense which counts objects below a threshold. As there exists a value of the CDF for every value of mass, there are infinite variations to the training data available to be generated from the simulation results. This continuous representation is particularly well-suited to neural network emulation, as the network can be trained and queried at arbitrary mass values and the traditional HMF $\mathrm{d}n/\mathrm{d}\ln M$ can be recovered by differentiating the predicted cumulative distribution.

\subsection{Network Architecture}\label{architecture}
The emulator is a fully-connected feed-forward neural network that takes cosmological and astrophysical parameters, mass, and redshift as inputs and outputs both a predicted halo count $N_p$ and a predicted model bias $\sigma_p$ learned through supervised training on simulation data. The neural network consists of seven hidden layers of width 40 with Tanh activation functions between each layer. The width and depth of the network were chosen through trial and error on a held-out validation parameter set: shallower and narrower networks led to under-fitting, while deeper and wider networks led to negligible decrease in cost. Tanh was selected over other activation functions because it resulted in the lowest model bias, most efficient training, and has the benefit of being continuously differentiable (unlike ReLU, PReLU, etc.) which is important when using results for forward modeling (see Eq.~\ref{eq:obsdist}). 

\subsection{Test and Training Data}
For each cosmology (i.e., unique combination of simulation input parameters) at each redshift, we randomly selected 15 halo masses from a log-uniform distribution and recorded the number of simulated halos above each mass as the training target, $N_s$. For training, all inputs (cosmological and astrophysical parameters, redshift, and mass) were linearly rescaled to a maximum value of 1, and the targets $N_{s}$ were rescaled as $\log_{10}(N_{s})/\max(\log_{10}(N_{s})$.

\subsection{Training}

\begin{figure}
    \centering
    \includegraphics[width =\textwidth]{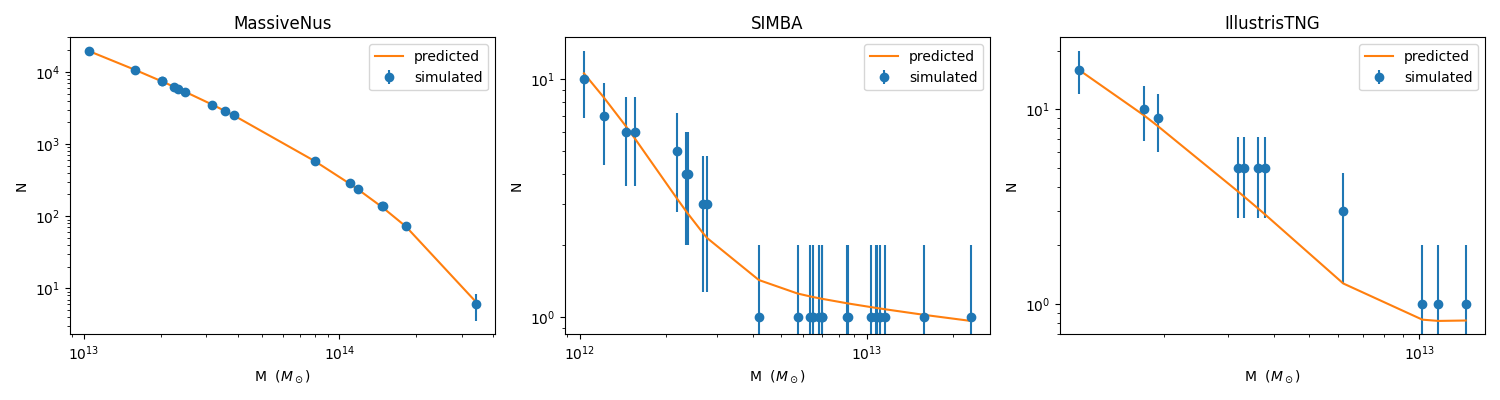}
    \caption{Examples of simulated and predicted cumulative mass count distributions for MassiveNus, SIMBA, and IllustrisTNG all at $z = 0$. The blue dots show the simulated number of objects with $m \geq M$ for a sample cosmology; the error bars represent shot noise. The orange lines show the continuous cumulative distributions predicted by a network trained on all cosmologies except the one plotted in blue. The cosmological parameters for these realizations are $M_\nu = 0.6$ eV, $\Omega_M = 0.3$, and $A_s = 2.1\times10^{-9}$ for MassiveNus and $\Omega_M = 0.119$, $\sigma_8 = 0.7722$, $\mathrm{SN}_1 = 0.1206$, $\mathrm{SN}_2 = 0.5732$, $\mathrm{AGN}_1 = 0.881$ and  $\mathrm{AGN}_2 = 0.1268$ for the CAMELS simulations. Because the CAMELS simulations have a significantly smaller box size, each count has a higher level of shot noise error. The higher variability in accuracy in the center and right plot is a result of statistical limitations, not model performance.}
    \label{fig:example}
\end{figure}

The networks were trained using 400 batches of 100 data points at each epoch. This means that 100 samples of the data were taken without replacement and each used to update network parameters, then all samples were replaced and this process was repeated 400 times per epoch. The training set was randomly shuffled such that the individual batches varied at each epoch. The Adam optimizer \citep{kingma_adam_2017} was used for gradient descent. Training was initiated with a learning rate (or gradient step size) of $5 \times 10^{-4}$ and a learning rate decay of 0.998 per epoch, then run for 3000 epochs. Figure~\ref{fig:example} shows an example of simulated vs. predicted counts at $z = 0$ for trained models in each simulation set. 

Learning rate and decay are highly sensitive parameters, especially with logarithmically scaled data. Decay rates smaller than 0.998 did not allow the network enough time between decays to make meaningful weight adjustments at a given step size, while larger decay rates made training slower with no added benefits. Smaller learning rates tended to land on local rather than global minima, while higher learning rates caused the initial training stages of the network to diverge. Due to the extreme logarithmic landscape of parameter space, one step in an inopportune direction can quickly lead the gradient into a region of parameter space that the optimizer cannot navigate away from, causing training to plateau. To remedy this, we developed a checkpoint algorithm. Every 10 epochs, the algorithm saves all the weights of the network. If the cost has increased more than $5\times$ since the last checkpoint, is NaN, or is infinite, it reverts to the previous checkpoint, clears all gradients, and continues. This safeguard is invoked very rarely and does not prevent the usual oscillation in cost, however, when invoked, it can allow an otherwise un-trainable iteration to continue training.

\subsection{Cost Function}
Because each truth value is a count of all objects above a certain mass, the random fluctuations in the truth values follow Poisson counting statistics (i.e., shot noise). This means we already know, analytically, how much of the discrepancy between a prediction and a simulated count should be attributed to sampling noise rather than genuine model error: for a Poisson process, the expected fluctuation in a count $N_s$ is $\sqrt{N_s}$. We therefore define a normalized bias term,

\begin{equation}
    B = \dfrac{|N_p - N_s|}{\sqrt{N_s}},
    \label{eq:bias}
\end{equation}
 which compares the difference between predicted counts $N_p$ and simulated counts $N_s$ to this expected noise level. If the network's error is dominated by statistical fluctuations noise rather than a systematic failure to learn the true underlying distribution, the difference between simulated and predicted counts $|N_s - N_p|$ should be within $\sqrt{N_s}$ ($1\sigma$) for the majority of predictions.

 $B$ can only be computed where a simulated truth value $N_s$ exists. When the network is applied to a cosmology that was never simulated, there is no $N_s$ to compare against, and no way to know how reliable a predicted value is. We therefore train the network to predict a second quantity, $\sigma_p$, alongside $N_p$ as a metric of reliability. The full cost function is

\begin{equation}
    C = B^2 + {(B - \sigma_p)}^2.
    \label{eq:cost}
\end{equation}
The first term trains the network to produce accurate counts $N_p$, penalized relative to the known shot-noise floor rather than in absolute terms, so the optimizer is not pushed to over-fit noise-dominated, low-count regions of mass/redshift space at the expense of higher-count regions. The second term trains $\sigma_p$ to track the realized value of $B$ for each training example, so that once training is complete, $\sigma_p$ can be evaluated for any input — including cosmologies outside the training set — and used as a confidence metric even when no ground truth is available. Cost is averaged over each batch and used to update gradients and weights.

\section{Results and Discussion}
\label{sec: Results}
\subsection{Model Bias}

\begin{figure}
    \centering
    \includegraphics[width=\textwidth]{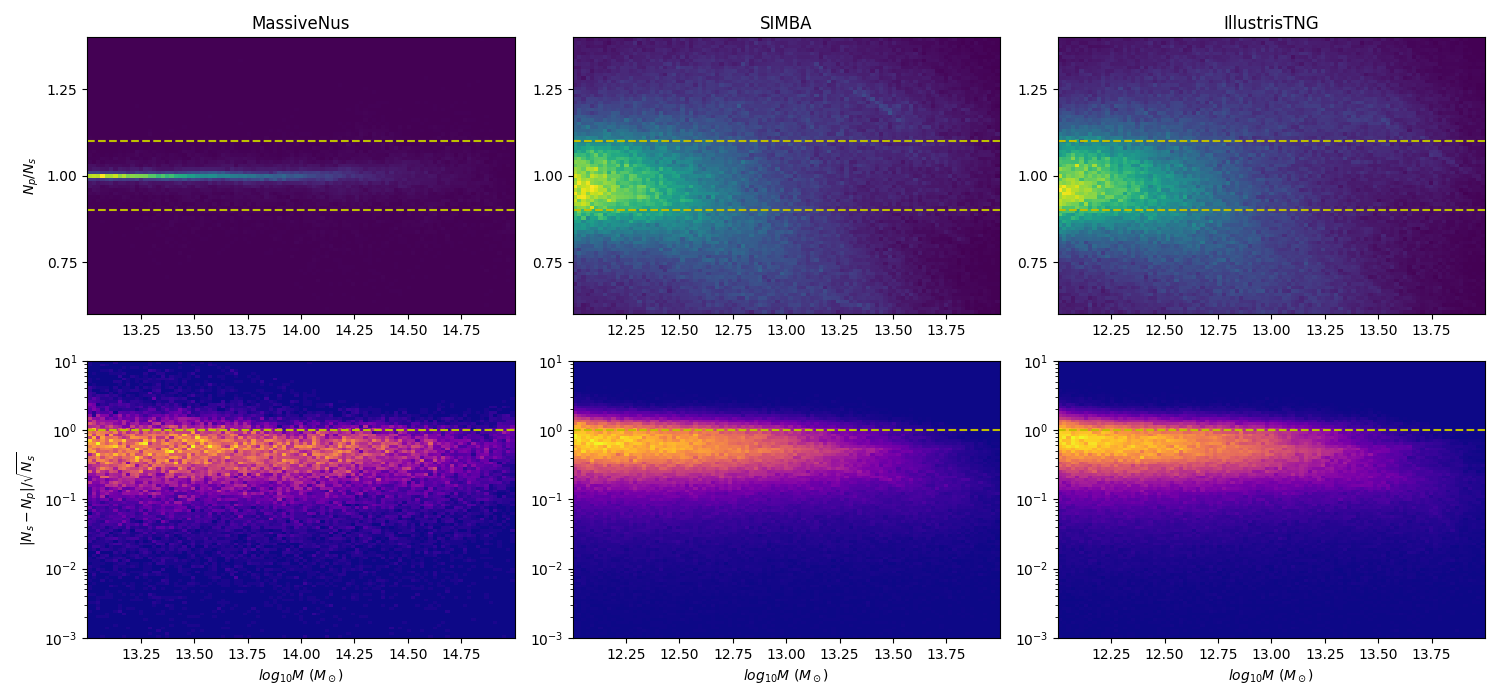}
    \caption{Model bias from leave-one-out tests across all simulations and all redshifts.  Top: histograms of the ratio of predicted to simulated counts $N_p/N_s$. Dashed lines indicated $\pm$ 10\% error. Bottom: histograms of model bias $|N_s-N_p|/\sqrt{N_s}$. Dashed lines show where the difference between simulated and predicted counts is equal to shot noise. This figure shows that fractional error does not take into account statistical limits when assessing the emulator's performance. Looking at fractional error alone might lead a reader to believe that the network performs significantly worse on CAMELS data. However model bias shows that this is not the case. The vast majority of leave-one-out tests have a model bias at or below shot noise, meaning the network is performing as well as it possibly can given the statistical constraints. }
    \label{fig:fracerrorhist}
\end{figure}

To assess how well the network can make predictions for cosmologies outside of its training set, leave-one-out (LOO) tests were performed for each simulation set. This was done by training the network on all but one simulated cosmology and then using data from the left-out cosmology as a test sample. \textit{All results presented are the application of the emulator to a data set not used in training}.
In the top panels of Figure~\ref{fig:fracerrorhist}, predicted counts $N_p$ from LOO tests as a fraction of simulated counts $N_s$ (fractional error) are histogrammed against cluster mass. 
The bottom panel shows model bias $|N_s-N_p|/\sqrt{N_s}$ from LOO tests for all cosmologies. In other works, fractional error is generally used to determine the accuracy of the model. However, it is important to note the difference between these two plots. When only looking at fractional error, it appears that results begin to diverge considerably for high masses. It also appears that accuracy is significantly better for MassiveNus than for the CAMELS suites (SIMBA and IllustrisTNG), whose smaller box sizes and correspondingly noisier counts widen the fractional-error distributions. However, when the difference between simulated and predicted values is compared to shot noise in the bottom panel, it becomes clear that the vast majority of predicted halo counts are within shot noise of simulated halo counts at all masses.

\begin{figure}
    \centering
    \includegraphics[width=\textwidth]{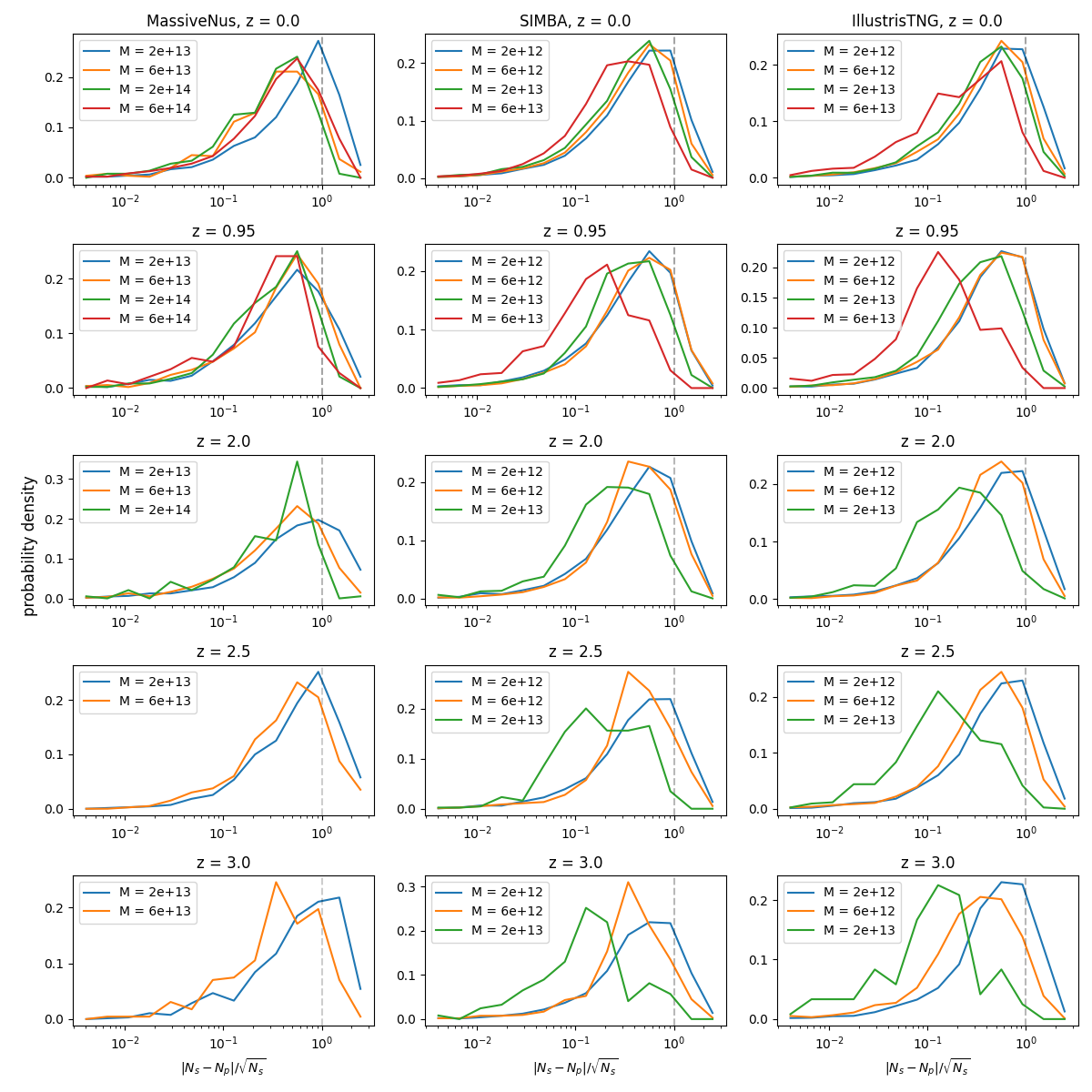}
    \caption{Model bias distributions $|N_s - N_p|/\sqrt{N_s}$ in a selection of mass (in units of $M_\odot$) and redshift bins for each simulation set. All data points at a given redshift in each simulation are binned by mass with bins log-centered at the masses shown on the legend. Bins containing less than 100 halos are excluded. All distributions peak at or below a bias value of 1, indicating that error at all redshifts and masses is dominated by statistical noise, not model bias.}
    \label{fig:linehists}
\end{figure}

The accuracy of the network's predictions with respect to shot noise is largely consistent across mass and redshift, though bias distributions tend to be lower and narrower at higher redshifts because there are fewer halo counts and therefore a smaller variance in halo counts. This is demonstrated in Figure~\ref{fig:linehists}, which shows the distribution of model bias for several mass and redshift bins. The distributions are generally centered around 1 (or slightly below), which indicates that on average, the difference between predicted and simulated counts is less than or equal to shot noise. Shot noise is an estimate of the average difference between some number of counts and the true, underlying mean of the distribution that value was pulled from. It is therefore expected that the model bias on some predictions exceeds shot noise. 

Table~\ref{tab:biasstats} summarizes the leave-one-out model bias across all mass and redshift bins for each simulation suite. Across all three suites combined, 87.72\% of the $1.2\times10^6$ individual LOO predictions have $B\le1$ and 99.93\% have $B\le3$, with a combined median bias of $B=0.42$. Statistically, it is expected that 99.73\% of measurements will fall within $3\sigma$ ($B \leq 3$) of the true model. This is true for both CAMELS simulations, however MassiveNus does not meet this criteria, meaning that some prediction errors are dominated by model bias. In the following section, we discuss how the predicted model error can be used to flag these high error predictions.

\begin{table}
    \centering
    \begin{tabular}{|c|c|c|c|c|}
        \hline
         Simulation & Median $B$ & Frac. $B\le1$ & Frac. $B\le3$& $N$ (LOO predictions) \\
         \hline
         MassiveNus & 0.40 & 88.93\%& 99.22\%& 55{,}098 \\
         \hline
         SIMBA & 0.42 & 87.56\%& 99.97\%& 570{,}922 \\
         \hline
         IllustrisTNG & 0.42 & 87.77\%& 99.97\%& 566{,}254 \\
         \hline
         All (combined) & 0.42 & 87.72\%& 99.93\%& 1{,}192{,}274 \\
         \hline
    \end{tabular}
    \caption{Summary of model bias $B = |N_s-N_p|/\sqrt{N_s}$ across all leave-one-out tests, all masses, and all redshifts, for each simulation suite. Fractions give the share of individual LOO predictions at or below the stated multiple of the shot-noise floor.}
    \label{tab:biasstats}
\end{table}

\subsection{Predicted Model Bias}
\begin{figure}
    \centering
    \includegraphics[width=1\linewidth]{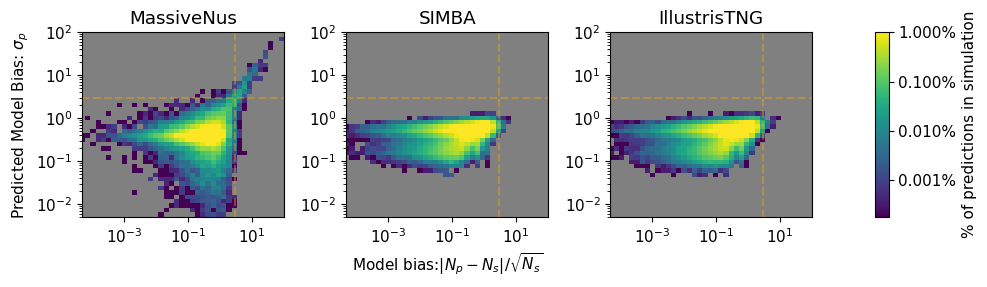}
    \caption{Two-dimensional histograms of model bias $B=|N_p-N_s|/\sqrt{N_s}$ (horizontal axis) against predicted model bias $\sigma_p$ (vertical axis) for every leave-one-out prediction in each simulation suite. Color shows the percentage of that suite's predictions in each bin, on a logarithmic scale. Dashed orange lines mark $3\times$ shot noise in each quantity, so predictions in the upper-right quadrant are high-bias predictions that $\sigma_p$ flags correctly. In MassiveNus, 0.78\% of predictions have $B>3$; $\sigma_p$ flags 76\% of these, and it tracks $B$ closely in the tail (all 65 predictions with $B>10$ have $\sigma_p>3$). In the CAMELS suites only 0.03\% of predictions exceed $B=3$, and $\sigma_p$ stays below $\approx1$ throughout, so no predictions are flagged.}
    \label{fig:histprederr}
\end{figure}

The predicted model bias $\sigma_p$ is the network's estimate of how much predicted halo counts deviate from the simulated values relative to shot noise. This is a built-in confidence metric, so that when applying the network to input parameters outside the training data, one can estimate the bias on a given prediction. It is a continuous function of input parameters. This value is a direct output of the network, produced alongside the value of predicted cluster counts. To assess the utility of the predicted model bias $\sigma_p$, we compare model bias to predicted model bias for each each prediction in each simulation in Figure~\ref{fig:histprederr}. 

As discussed above, shot noise is an estimate of the average difference between some number of counts and the true, underlying mean of the distribution that value was pulled from. It is therefore not expected that the model bias on some predictions exceeds shot noise. The predicted model error is a measure of how the difference between the true underlying value and the model predictions compare to shot noise. As shown in the second and third panels of Figure \ref{fig:histprederr} as well as in Table \ref{tab:biasstats}, model bias rarely exceeds shot noise in the CAMELS simulations, and it is therefore unsurprising that that the predicted model error never exceeds 1. In the MassiveNus simulations, shown in panel 1 of Figure \ref{fig:histprederr}, we see that there are in fact a sizable number of predictions with model error as high as $10^2$, and that the predicted model error effectively flags these predictions as being unreliable.  Of the predictions with $B>3$ in the MassiveNus simulation, the model flags 76\% as having $\sigma_p > 3$. The remaining 24\% (0.19\% of all MassiveNus predictions) are comparable to the $\approx0.27\%$ expected from shot noise alone. 

\subsection{Model Comparisons}

\begin{figure}
    \centering
    \includegraphics[width = \textwidth]{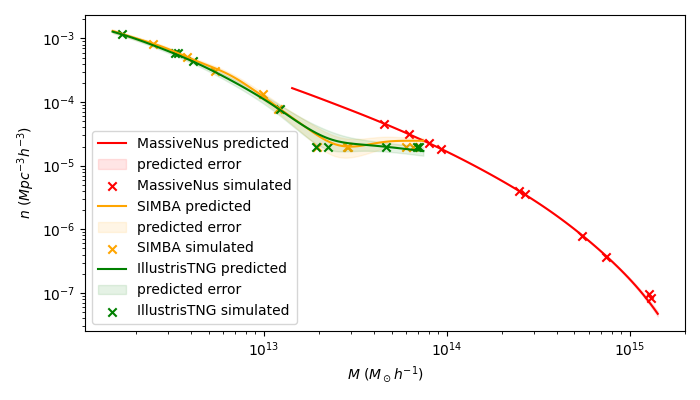}
    \caption{Side by side comparison of predicted and simulated mass distributions using fiducial values for all simulation sets (see Tables~\ref{tab:CAMELSfid} and~\ref{tab:MNusfid}). The solid lines and error bars are mass distributions and biases predicted by the trained neural network. The $x$'s are simulated values. The flattening of the SIMBA and IllustrisTNG distributions at higher redshifts is likely due to the suppression of high mass cluster formation in simulations with smaller box sizes.}
    \label{fig:sim_comp}
\end{figure}

CAMELS and MassiveNus simulations span different mass ranges, making a direct comparison informative. Figure~\ref{fig:sim_comp} shows the simulated and predicted results using fiducial values for all three simulations. The general trends of the distributions agree quite well, however there are some discrepancies where the two mass ranges overlap. Each simulation suite is a single finite-volume realization, so beyond the Poisson shot noise already captured by our bias metric, cross-suite comparisons are also subject to sample variance from fluctuations in the large-scale density mode sampled by each box \citep{hu_sample_2003}. While these effects are generally small for MassiveNus's $(512/h\,\mathrm{Mpc})^3$ box, in the CAMELS suites --- whose $(25/h\,\mathrm{Mpc})^3$ boxes are more than 20 times smaller in linear size --- the already rare high-mass halos are sparsely sampled to begin with, and sample variance can suppress their count further, producing the rapid flattening of the distribution when approaching the mass domain of MassiveNus. However, the difference in predicted counts in this regime is at most a few halos.

\begin{figure}
    \centering
    \includegraphics[width =\textwidth]{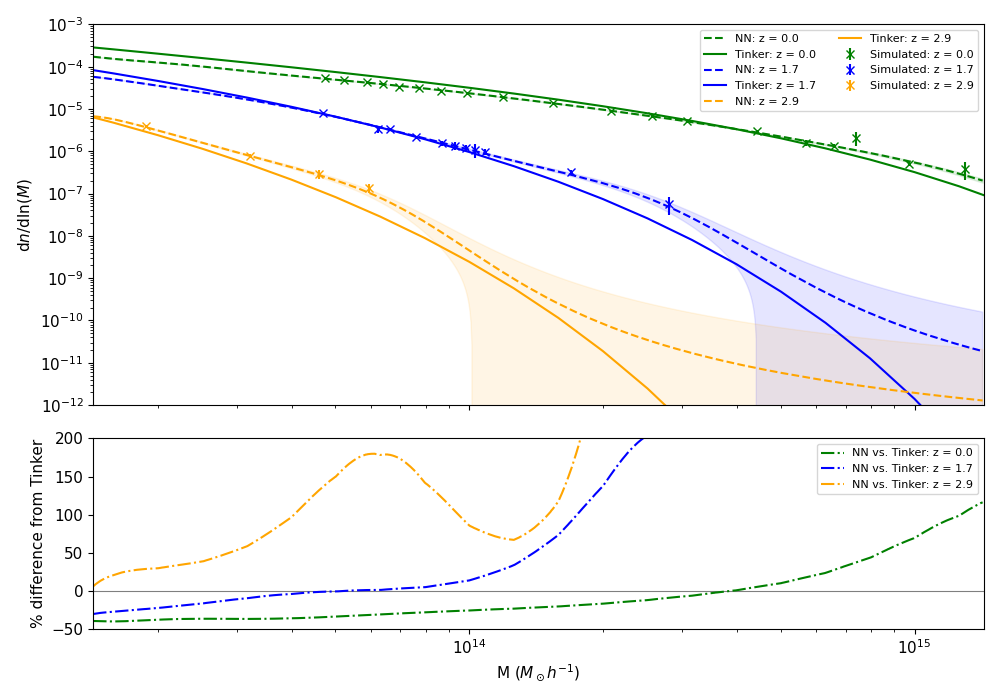}
 \caption{Comparison between the \citet{tinker_toward_2008} HMF (Colossus, \citealt{diemer_colossus_2018}) and an HMF generated by our emulator trained on the MassiveNus simulations. HMFs are shown at at $z=0$, $1.7$, and $2.9$ using $H_0=70$ km/s/Mpc, $n_s=0.97$, $\Omega_b=0.046$, $\Omega_M=0.3$, $\Omega_\Lambda=0.7$, $M_\nu=0$, and $A_s=2.1\times10^{-9}$ ($\sigma_8=0.85$, computed from $A_s$ with CAMB). \textit{Top:} dashed lines show the emulator prediction from a network trained on all MassiveNus cosmologies except the one shown; solid lines show the Tinker HMF. Crosses are simulated values of $\mathrm{d}n/\mathrm{d}\ln M$ from the MassiveNus simulation, obtained from finite differences of the simulated cumulative counts between adjacent sampled masses, with Poisson errors on the binned counts, $\sqrt{\Delta N}/(V\,\Delta\ln M)$; their fractional errors grow with mass and redshift as halos become rarer. \textit{Bottom:} signed percent difference $(\mathrm{NN}-\mathrm{Tinker})/\mathrm{Tinker}$. Beyond the most massive simulated halo at each redshift the comparison is not constrained by data, and the $z=2.9$ curve lies outside the range ($z\le2.5$) over which the Tinker redshift evolution was calibrated. Similar comparisons of emulators to the Tinker fit appear in \citet{mcclintock_aemulus_2019, nishimichi_dark_2019, bocquet_mira-titan_2020}.}
    \label{fig:tinker}
\end{figure}

In Figure \ref{fig:tinker}, we compare the results from these models to HMFs from \citet{tinker_toward_2008} generated by the Colossus code \citep{diemer_colossus_2018} for several redshifts at the fiducial values used in MassiveNus. We restrict this comparison to MassiveNus because the \citet{tinker_toward_2008} ansatz is calibrated on dark-matter-only N-body simulations and does not account for the baryonic feedback processes present in the CAMELS suites, making it an unsuitable benchmark for SIMBA and IllustrisTNG For low redshifts, the results of the neural network agree quite well with the Tinker predictions. At higher redshifts, there are larger discrepancies but generally within the range of statistical fluctuations. Part of this discrepancy reflects the sample variance discussed above. The logarithmic scaling of outputs means the network is never trained on zero truth values, resulting in a slight systematic over-prediction for masses at the high end of the training data.  As demonstrated in the figure, even where the shape of the predicted HMF is not quite in agreement with the Tinker predictions, it still follows the simulated results quite well. 

The residuals exhibit some mass dependence, with opposite signs at the two ends. At low masses the network lies below the Tinker fit; this is consistent with a mismatch in halo mass definition: \citet{tinker_toward_2008} calibrate their fit to the mean-overdensity mass $M_{200m}$, while we evaluate it at $M_{200c}$, the critical-overdensity definition matching our simulated halo catalogs. \citet{bocquet_halo_2016} show that halo abundances computed under mean- and critical-overdensity definitions diverge most strongly at low mass, and \citet{bocquet_mira-titan_2020} similarly find that the Tinker fit overestimates halo abundance by up to $\sim$10\% below $10^{14}\,h^{-1}M_\odot$ under a mismatched definition -- consistent with the sign and location of the low-mass offset we find. In the exponential tail, by contrast, the simulated abundances lie above the Tinker curve, and the difference grows with mass and redshift. \citet{nishimichi_dark_2019} find a comparable mass-dependent bias toward the massive end at $z\gtrsim1$, which they attribute to the initial conditions of the simulations used to calibrate the Tinker fit, and \citet{tinker_toward_2008} themselves note that such transients most strongly affect the rarest peaks.

This growth with redshift is consistent with the limits of the Tinker fit's own calibration. \citet{tinker_toward_2008} calibrate the redshift evolution of their fit only up to $z=2.5$; our $z=2.9$ curve therefore lies outside the range constrained by their simulation data, and beyond the last simulated halo in each panel the comparison is between two extrapolations. Other comparisons of emulated or simulated HMFs against the Tinker fit report a similar growth in the discrepancy with redshift: \citet{mcclintock_aemulus_2019} found a mean absolute residual of 3.4\% (consistent with $\sim$5\% at $z=0$) growing to over 100\% at $z=2$--$3$, and \citet{nishimichi_dark_2019} likewise found deviations from Tinker that grow with redshift. The growth of our discrepancy with redshift is therefore qualitatively consistent with the known limitations of the Tinker fit, in addition to the sample-variance and log-scaling effects discussed above.

\section{Conclusion and Future Work}
\label{sec: Discussion and Conclusion}

In this work, we have presented a machine learning-based approach to emulating halo mass distributions and subsequently the halo mass function (HMF) for cosmological simulations. This method utilizes a neural network to predict the number density of galaxy clusters as a function of mass, based on cosmological and astrophysical parameters. We have demonstrated the accuracy and generalizability of our approach by applying it to multiple simulation sets and comparing the results to analytical HMF predictions. The flexibility of this method, which captures the intricate behavior of the HMF without relying on parametric models or data reduction techniques, enables simulations using a wide range of cosmological and astrophysical parameters to be efficiently expanded via emulation. The method requires no simulation-specific tuning beyond a configuration file specifying the relevant input parameters and data columns, so applying it to a new simulation suite is straightforward; we publicly release our implementation, including this configuration interface, at \url{https://github.com/oross314/HMF_Modeling}.

The network outputs a self-calibrating uncertainty estimate $\sigma_p$ alongside each predicted halo count, providing a built-in confidence metric for emulated cosmologies outside the training set. It is important to note that the resulting HMFs are extensions of the simulated data and inherit the assumptions and errors contained therein. As with any data-driven emulator, predictions are most reliable within the range of parameters spanned by the training simulations — extrapolating beyond it is inherently less certain, a limitation the predicted bias is designed to flag.  This neural network-based method preserves fine-scale fluctuations and captures the full complexity of the HMF. We have demonstrated that $\sigma_p$ correctly flags regimes where model bias exceeds reasonable statistical fluctuations in halo counts.

The emulated HMFs can be employed in various applications, most notably forward modeling to predict observable distributions. By converting the distribution of halo masses to observables, predictions can be directly compared with observations, enabling constraints to be placed on cosmological and astrophysical parameters. 

\bibliography{HMF}
\bibliographystyle{aasjournalv7}

\end{document}